# THE FIVE-HUNDRED-METER APERTURE SPHERICAL RADIO TELESCOPE (FAST) PROJECT

RENDONG NAN[1,2,4], DI LI[1,3,5], CHENGJIN JIN[1], QIMING WANG[1], LICHUN ZHU[1], WENBAI ZHU[1], HAIYAN ZHANG[1,2], YOULING YUE[1] AND LEI QIAN[1]

1. National Astronomical Observatories,

   Chinese Academy of Sciences,

   A20 Datun Road, Chaoyang District,

   Beijing 100012, China.

2. Key Laboratory for Radio Astronomy,

   Chinese Academy of Sciences,

   Nanjing 210008, China

3. Jet Propulsion Laboratory,

   California Institute of Technology,

   Pasadena, CA 91109, USA

4. nrd@bao.ac.cn

5. dili@nao.cas.cn

Five-hundred-meter Aperture Spherical radio Telescope (FAST) is a Chinese mega-science project to build the largest single dish radio telescope in the world. Its innovative engineering concept and design pave a new road to realize a huge single dish in the most effective way. FAST also represents Chinese contribution in the international efforts to build the square kilometer array (SKA). Being the most sensitive single dish radio telescope, FAST will enable astronomers to jump-start many science goals, for example, surveying the neutral hydrogen in the Milky Way and other galaxies, detecting faint pulsars, looking for the first shining stars, hearing the possible signals from other civilizations, etc.

The idea of sitting a large spherical dish in a karst depression is rooted in Arecibo telescope. FAST is an Arecibo-type antenna with three outstanding aspects: the karst depression used as the site, which is large to host the 500-meter telescope and deep to allow a zenith angle of 40 degrees; the active main reflector correcting for spherical aberration on the ground to achieve a full polarization and a wide band without involving complex feed systems; and the light-weight feed cabin driven by cables and servomechanism plus a parallel robot as a secondary adjustable system to move with high precision. The feasibility studies for FAST have been carried out for 14 years, supported by Chinese and world astronomical communities. Funding for FAST has been approved by the National Development and Reform Commission in July of 2007 with a capital budget ~ 700 million RMB. The project time is 5.5 years from the commencement of work in March of 2011 and the first light is expected to be in 2016.

This review intends to introduce FAST project with emphasis on the recent progress since 2006. In this paper, the subsystems of FAST are described in modest details followed by discussions of the fundamental science goals and examples of early science projects.

1. Introduction

The Large Telescope (LT), referred to as the Square Kilometre Array (SKA) nowadays, was proposed by astronomers from 10 countries including China at the General Assembly of the International Union of Radio Science (URSI) in 1993, which started a worldwide effort to develop the scientific goals and technical specifications for a next generation radio telescope. Different technological solutions have been put forward and studied by institutes participating in the SKA. FAST, which is to be built in a karst depression in Guizhou province of southwest China, is the Chinese engineering concept proposed and extensively investigated since 1994 for realizing the SKA units. Cooperating with Chinese and international astronomical communities, FAST team of the National Astronomical Observatories of the Chinese Academy of Sciences (NAOC) has successfully conducted studies on the critical technologies including the site surveying, an active reflector, a light-weight feed support system, remote and accurate measurements and control, and receivers[1].

An international review and advisory conference on science and technology of FAST was held in Beijing in March 2006. The review panel unanimously concluded that the FAST Project is feasible and recommended that the project moves forward to the next phase of detailed design and construction as soon as possible. Funding for project FAST has finally been approved by the National Development and Reform Commission (NDRC) in July of 2007. The approved budget is now 700 million RMB. At the end of 2008, the foundation has been laid. The construction period is 5.5 years from the commencement of work in March 2011. The first light is expected to be in 2016.

In a 2006 review paper [1], the general technical specifications of the subsystems of FAST have been briefly discussed. Since that was before the approval of the FAST proposal, much of the subsystem concept was in a crude form. It is necessary to update the community of the status of the FAST project and progresses in the system designs. This paper is intended to serve as an overview of the approved FAST project. In section 2, the general technical specifications and scientific motivations of FAST are given. The subsystems of FAST are introduced in detail in section 3, with the requirements on them from the FAST science projects presented, too. In section 4, the FAST sciences goals are discussed.

## 2. General Technical Specification and Scientific Motivation

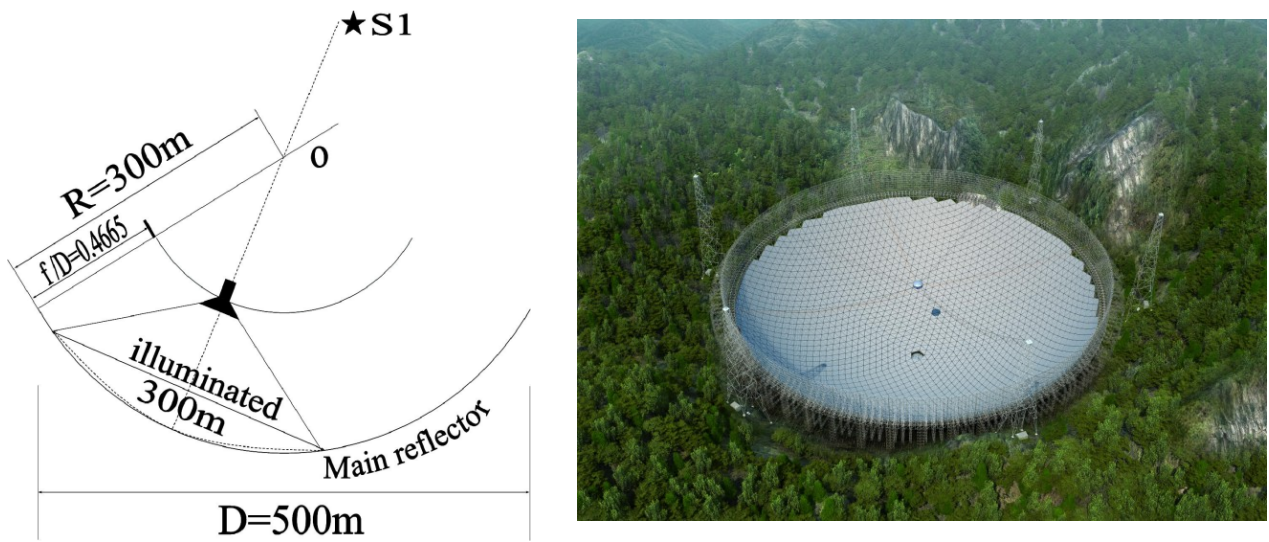

Figure 1: Left: FAST optical geometry, right: FAST 3-D model

FAST is an Arecibo-type spherical telescope. Figure 1 illustrates the optical geometry of FAST and its three outstanding features: the large karst depressions found in south Guizhou province as the sites [2], the active main reflector of 500 m aperture which directly corrects for spherical aberration [3], and the light-weight feed cabin driven by cables and a servomechanism [4] plus a parallel robot as a secondary adjustable system to carry the most precise parts of the receivers. Inside the cabin, multi-beam and multi-band receivers will be installed, covering a frequency range of 70MHz - 3 GHz. The telescope will be equipped with a variety of instruments and terminals for different scientific purposes. The main technical specifications of FAST are listed in Table 1.

Table 1 FAST main technical specification

Spherical reflector: Radius=300m, Aperture=500m, Opening angle=112.8°

Illuminated aperture: $D_{ill}$ =300m

Focal ratio: f/D =0.4665

Sky coverage: zenith angle 40°, tracking time ~ 6h

Frequency: 70 MHz - 3 GHz (up to 8GHz, future upgrading)

Sensitivity (L-Band): A/T ~ 2000, system temperature $T_{sys}$ ~ 20 K

Resolution (L-Band): 2.9′

Multi-beam (L-Band): beam number=19 (future focal plane array >100)

Slewing time: <10 minutes

Pointing accuracy: 8″

Deep depression and feed cabin suspension system enable FAST reflector a large opening angle, covering a large zenith angle up to 40° with full illuminated area of a diameter of 300 m. Larger zenith angle up to 60°, which will extend the sky-coverage of FAST south beyond the galactic center, is feasible by using some particular feeding technologies, such as Phased Array Feed (PAF), which is to be employed at the focus in the future. The raw sensitivity in L-band, the core band for most significant sciences of FAST, reaches 2000 $m^2$ $K^{-1}$, thanks to the huge collecting area and up-to-date receiving system. Nineteen beams of horn-based receivers in the band are planned to increase the survey speed. Maximum slewing time is set to 10 minutes which is restricted by the power of high-voltage electromotor.

The coverage of S-band and below meets the requirements by the dominant part of science cases of FAST. 3GHz is set as the frequency upper limit of the first phase of the telescope construction, which would reduce the control accuracy requirements by a factor of 2 compared with the higher band limit, 8 GHz, specified in the preliminary design of the year 2000. Division of two construction phases will largely reduce construction venture, project time and capital budget.

The fundamental scientific motivation for building the largest radio telescope is to survey the radio universe by utilizing FAST's unparalleled sensitivity and high surveying speed. FAST should help to enhance our understanding of cosmology, galaxy evolution, interstellar medium (ISM) life cycle, star formation and exoplanets. The scientific goals include

- Survey the Galactic ISM in HI at a resolution comparable to the current large scale CO surveys;
- Discover ~4000 new Galactic pulsars and search for the first extragalactic pulsars;
- Detect tens of thousands of HI galaxies and detect individual massive galaxies up to z~1;
- Spectroscopic survey of the radio spectra of rich Galactic sources with continuous coverage between 70 MHz and 3 GHz;
- Search for radio signals from exoplanets.

In section 4, these goals will be discussed in more detail and examples of early science projects will be presented.

## 3. Critical Technologies and Subsystems

The telescope engineering can be divided into 6 major subsystems, including site surveying, exploration, and earthwork in the selected site; main active reflector; feed cabin suspension; measurement and control; receivers; and observatory. Figure 2 illustrates the subsystem composition of the FAST.

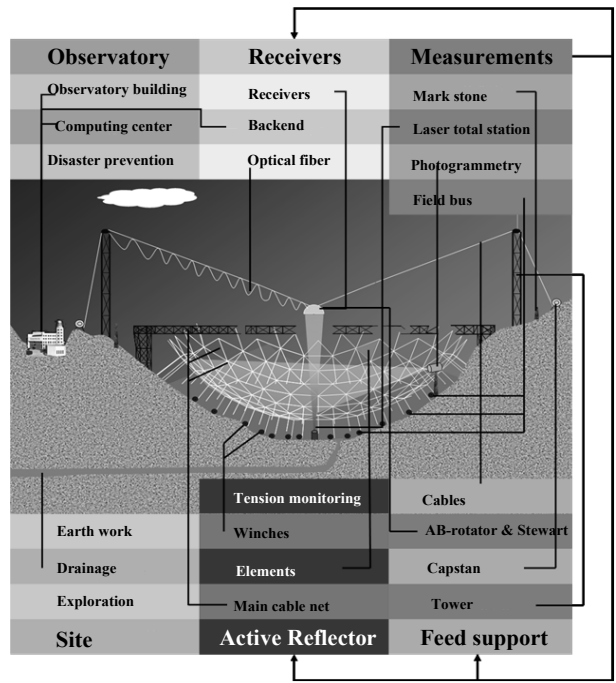

Figure 2 Telescope composition

A feasibility study of critical technologies started in the NAOC since 1994. More than one hundred scientists and engineers from 20 institutions are involved in the joint research project. Critical technologies to be employed in the construction have been certified and will be discussed below. We will discuss the five subsystems in this paper, namely site, active reflector, feed support, receivers, measurements and control. The observatory subsystem, which has not been studied in detail, will be discussed in future papers.

**3.1. Site surveying**

**3.1.1 The site**

A practical way to build a large spherical telescope is to make extensive use of existing depressions which are usually found in karst regions. Site surveying in Guizhou province started in 1994, including geo-morphological features and the distribution of the karst depressions, climate, engineering environment, social environment, and radio interference. At least 400 candidate depressions were investigated with remote sensing, and the Geographic Information System. The expense of earthwork largely depends on the geometrical profile of a depression.

The Dawodang depression in south Guizhou has been selected as the telescope site (Figure 3). Total earthwork is estimated to be ~ 1,000,000m$^3$ according to the comparison of the reflector model with the digital terrain model images of the depression. The relatively low latitude (~26°N) of the site enables the observation of more southern galactic objects. The mild climate of the subtropical zone with a few days of frost and snow without ice build-up enable survival of low cost structures. There is no inundation of karst depressions because of their good drainage, but a tunnel is still budgeted in order to ensure telescope safety. No devastating earthquake has ever been recorded in history. The remoteness and sparse population guarantee a clean RFI environment and the safety of future FAST instruments. An agreement on a temporary radio quiet zone around the site has already been signed by the Chinese Academy of Sciences (CAS) and Guizhou provincial government.

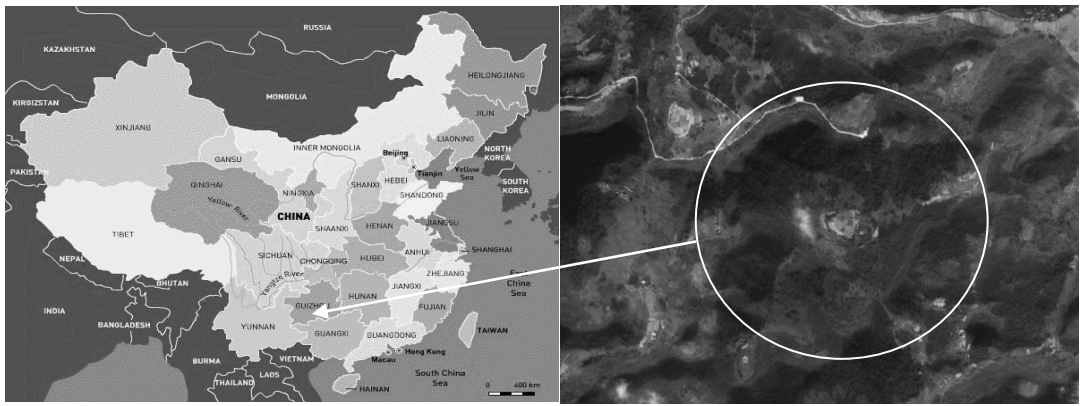

Figure 3 Left: Depression Dawodang, East: 107°21′ North: 25°48′, Altitude: ~1000m.
Right: image by quick bird with resolution of 0.6 m, the dimension of the circle is ~1000m.

Starting from January of 2010, a detailed FAST site exploration has been completed. High accuracy topographic map of FAST site has been measured, as shown in Figure 4. The RMS error is about 0.2m.

### 3.1.2 Requirement from FAST sciences

Radio observations are often affected by radio frequency interference (RFI). To realize the high sensitivity of FAST, the radio environment of FAST should be very quiet. Dawodang has not only the karst depression with size and shape suit for the construction of FAST, but also a very quiet radio environment. The surrounding mountains provide good shielding against RFI. Guizhou resides on a geological stable tectonic plate with very few earthquakes. The climate there is also moderate. All these make the Dawodang site ideal for radio sciences of FAST.

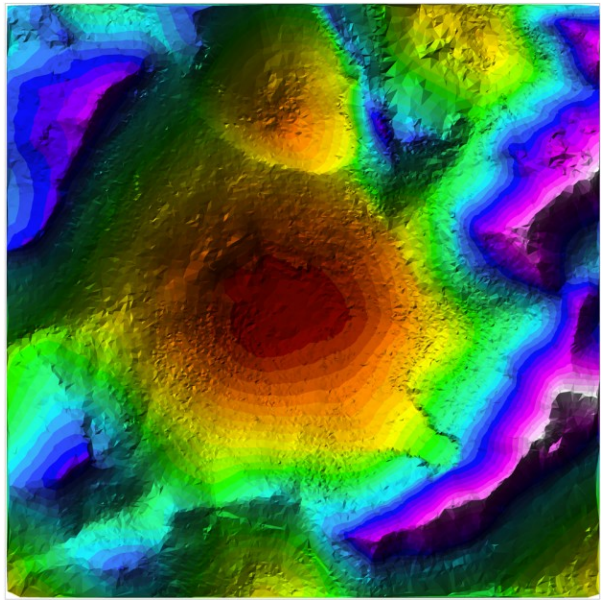

Figure 4 High accuracy (~0.2m) topographic map of FAST site.

### 3.2. Primary Reflector and Supporting Structures

### 3.2.1. Main active reflector

As a huge scientific device, the supporting structure of the radio telescope FAST demands special requirements beyond those of conventional structures. The most prominent one is that the supporting structure should enable the surface formation of a paraboloid from a sphere in real time through active control. Fortunately, the peak deviation of the paraboloid of revolution from the spherical surface is only about 0.67m [5] across the illuminated aperture of about 300 m. An adaptive cable-net design [6] has been proposed for two main reasons: first, the small difference required by the deformation mentioned above should be easily achieved within the elastic limit of ordinary cable wires; second, the cable-net structure should accommodate with the complex topography of karst terrain easily, which will avoid heavy civil engineering between actuators and the ground.

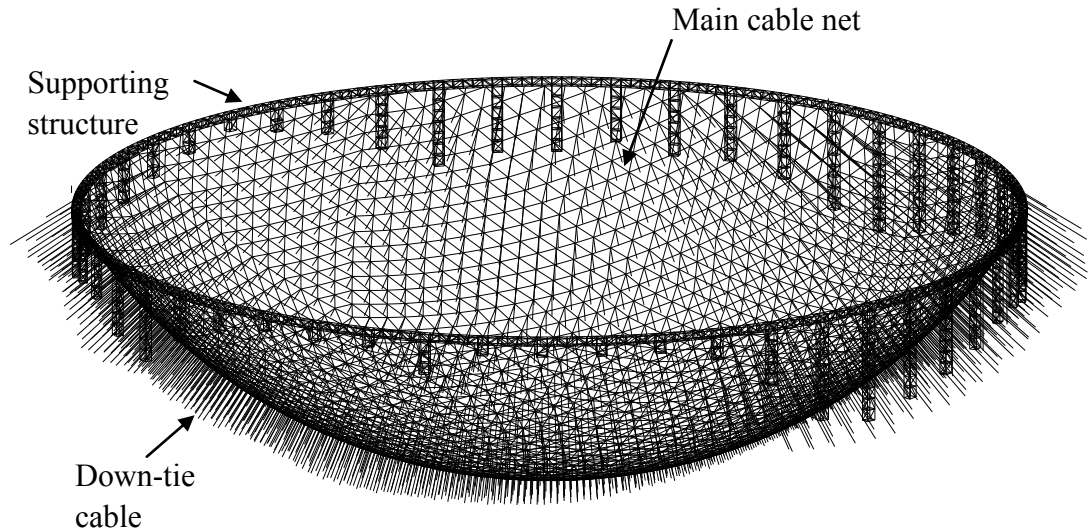

Figure 5 Concept of the adaptive cable-net structure, the supporting structure for the FAST reflector

### 3.2.2 Cable-net structure

The initial design of the supporting structure has been completed. The outer edge of the cable-net is to be suspended from the steel ring beam of a diameter of 500 m. The cable-net consists of two sets of cables in orthogonal directions (Figure 5). The aluminum panels are attached to the cable-net through adaptive connectors allowing compensation for small changes in length and angle during surface deformation. The crossed nodes of the cable-net are used as control points, which are tied to the control actuator by down tied cables. By controlling the actuator using the feedback from the measurement and control system, position of these control points can be adjusted to form the illuminated aperture of diameter 300 m in different regions of the cable-net. Figure 6 shows a close-up look of one cable-net node.

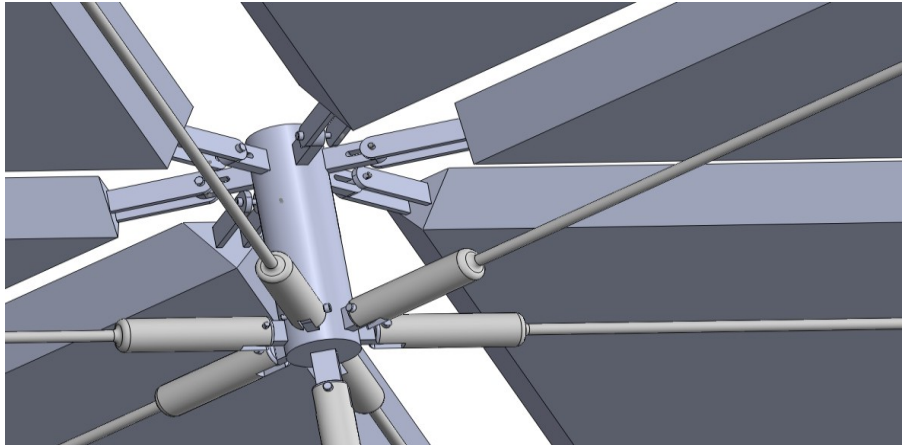

Figure 6 The node of the cable net of the back frame. In the lower part of the node, there are six cables of the back frame and a down-tied cable. In the upper part, the elements of the main reflector are connected to the node by adaptive junctions.

Extensive numerical study has been performed to investigate the feasibility of this design. Among six types of spherical cable-net patterns investigated, the geodesic trianglular element is selected, which have most uniform tension in all cables and least panel element types of 187. There are about 4400 triangular elements on the surface in total supported by the cable-net consisting of about 7000 steel cables. This results in about 2300 nodes and corresponding driving cables down-tied to actuators on the ground. All driving cables of the spherical surface are independent from each other, which makes it possible to adjust the cross section of each cable according to the load. The deformation strategy of the internal working part of the reflector is directly related to the internal forces of the cable-net and the actuator stroke, both of which are directly related to the cost of FAST. Numerical analysis has been performed to compare the three kinds of deformation strategies, whose relative position to the spherical surface are shown in Figure 7. The results show that parabola III has better performance in both internal forces of cable-net and actuator stroke.

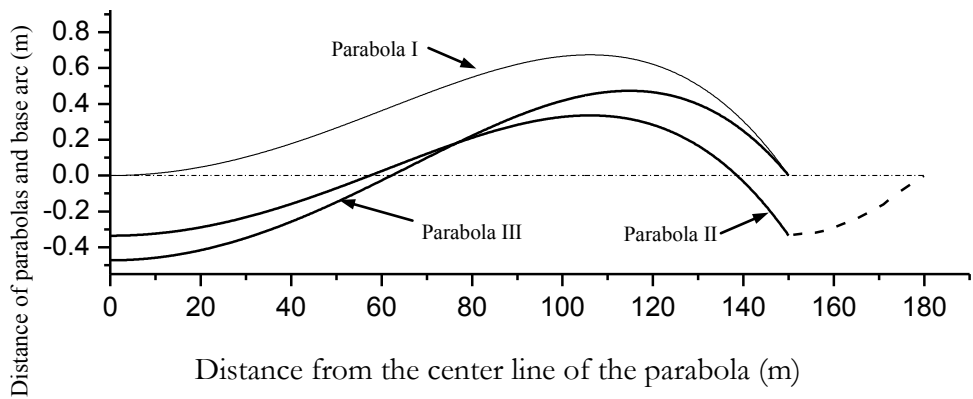

Figure 7 Relative position between base arc and parabolas of three deformation strategies.

### 3.2.3 Back frame

A reflector element is consisted of back frame, adjusting bolts, panels and purlins. Based on the material used, the back frames can be divided into two main categories: aluminum structure and steel structure. Purlins and panels are made of aluminum. The surface of each panel unit has been adjusted to the spherical surface of radius 318.5m other than the overall 300m radius of the main reflector, which will make it better fit to a paraboloid [7]. The adjusting bolt is made of stainless steel to avoid electrochemical corrosion. For the steel structure, four kinds of back frames have been developed (Figure 8): single-layer structure, suspended structure, double-layer structure, and space bolt-ball net frame structure.

Of the four structures, the space bolt-ball net frame structure has the best stiffness. Moreover, it is simple to manufacture and has high assembly efficiency. A new tri-pyramid grid structure for the bolt-ball net frame is developed with aluminum tube-type rods, bolts ball, purlins and panels (see Figure 9), except for the stainless steel connecting bolt. This structure is light and corrosion proof, both critical for the overall optimization of the reflector. Therefore, the space bolt-ball net frame structure is the first choice in the construction of the reflector.

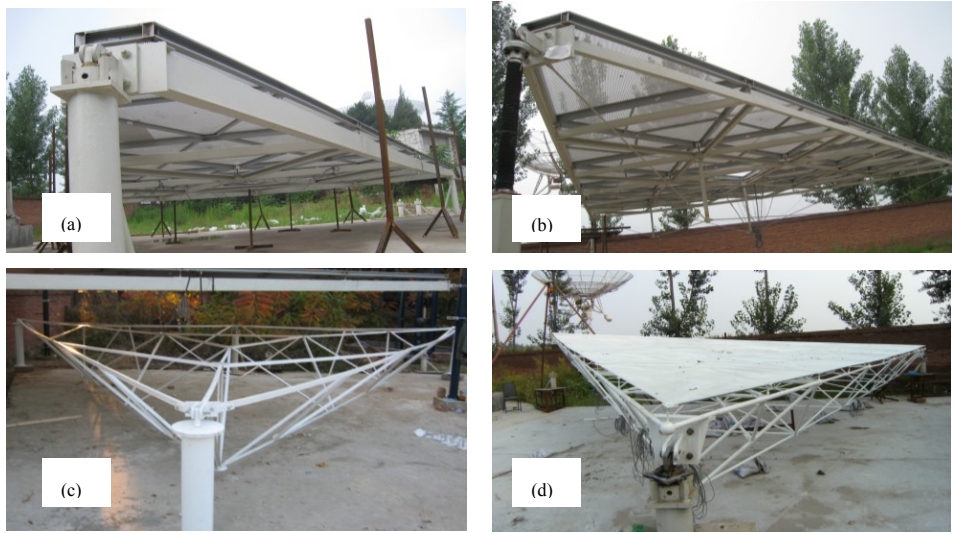

Figure 8 Prototypes of four kinds of back frames: (a) single-layer structure, (b) suspended structure, (c) double-layer structure, (d) space bolt-ball net frame structure.

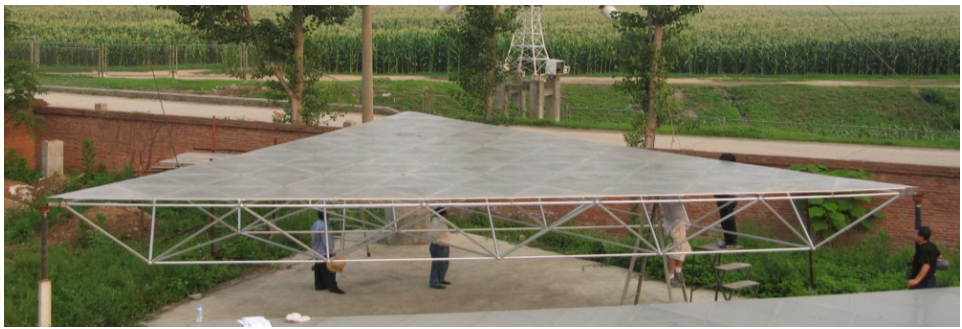

Figure 9 Prototype of the aluminum back frame.

### 3.2.4 CFD simulation of the wind around the FAST

Several environmental factors, including the wind force and temperature variation can cause deformation of FAST main active reflector, which is supported by a flexible cable-net system. The special terrain of the karst depression leads to the complex wind flowing pattern around FAST. The influence of the wind on the main active reflector is studied via numerical simulation. An example of unstructured mesh near the site is shown in Figure 10.

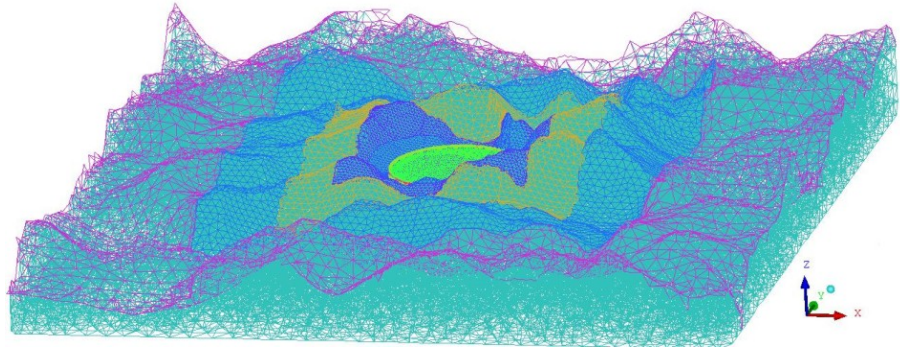

Figure 10 The unstructured mesh near the terrains

Based on the computational fluid dynamics (CFD) model, parameters such as pressure coefficient, pressure distribution and velocity distribution can be obtained for further design and construction of the structure.

### 3.2.5. Requirement from FAST sciences

FAST sciences put stringent limits on surface deformation of the main active reflector. The position of all the nodes need to be controlled precisely, which depends not only on the structure of the cable net, the back frame and the reflector elements, but also on status of thousands of down-tied cables and actuators. The tracking of FAST is realized by adjusting the reflector in real-time. So the speed of the deformation, which is related to the speed of actuators, should be high enough (about 15° per hour in the extreme case) to track a target in the sky. The shape and size of the panels are carefully chosen to minimize the required deformation from a sphere to form a paraboloid and to reduce intrinsic polarization.

### 3.3. Cabin Suspension

There is no solid connection between the reflector and the feed cabin because of the large dimension of the telescope. If Arecibo design were adopted, there would be 10,000 tons of metals hang in the air above the reflector, which is impractical. Instead, the feed cabin of FAST is supported and driven by cables and servomechanism. Inside of the cabin, secondary adjustable system is employed to achieve the required accuracy.

The designs consist of three essential parts including the cable network that supports and drives the feed cabin, the secondary adjustable devices inside the cabin that carry the most precise part of the receivers and the close loop control. The sketch of this mechanism-electronics-optics integrated design is shown by Figure 11.

Numerous models were made to test complex issues in the cabin suspension subsystem. Because of the large dimension and the complicated dynamics of cabin suspension, similarity laws may not guarantee the accuracy of down-scaled models. According to the recommendations of the international review panel, FAST team has carried out end-to-end simulation cooperated with MT Mechatronics and Technical University Darmstadt. This analysis has confirmed the feasibility of the concept and yielded critical results for system optimization. The displacement of feed cabin after first adjustment control can be constrained to be within a few centimeters. The achievable accuracy of feed position with the help of the secondary stabilizer turns out to be a few millimeters, which meets the requirement of telescope pointing [8].

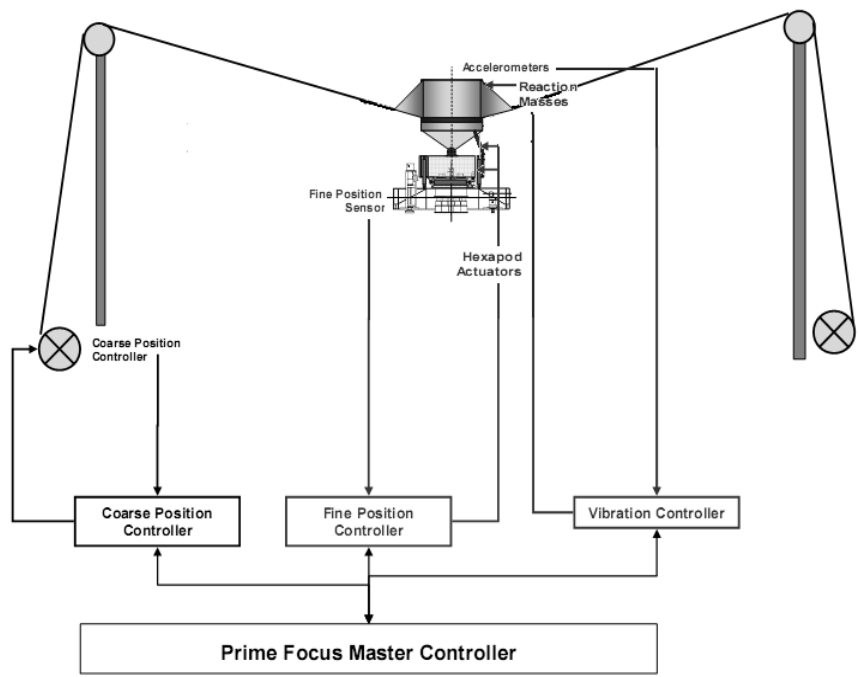

Figure 11 Sketch of the cabin suspension system.

### 3.3.1 Main position control system

The main position control system moves the feed cabin with suspension cables. Six suspension cables are driven and supported by six towers and capstans which are uniformly distributed on a circle with a diameter of 600m. It acts as the preliminary adjustor to keep control error of the cabin position within 100mm. The position error of the receiver in the cabin is further corrected by the fine drive system. Mechanically the main position control system is a 6-cable-driven parallel manipulator, where the cabin is the movable platform.

As the cabin moves slowly in the focal surface, the 6 cable forces and the natural tilt of the cabin vary accordingly. An equilibrium analysis of the cabin-cable suspension system based on cable force optimization shows that these variations depend on the cable-driven parallel manipulator and the focal surface. Figure 12(a) shows the variation of the cabin tilt on the whole focal plane. Fortunately the optimal natural tilt is almost always along the radial direction and favors the desired orientation. This fact provides a possibility to make full use of this natural tilt for minimizing the pose control task of the receiver platform in the cabin. Figure 12(b) shows the tension force variation of one suspension cable in the focal surface. Due to axial symmetry, similar force variations are expected for other cables after 60 degree rotation. For each suspension cable, the force ranges from 120kN to 280kN.

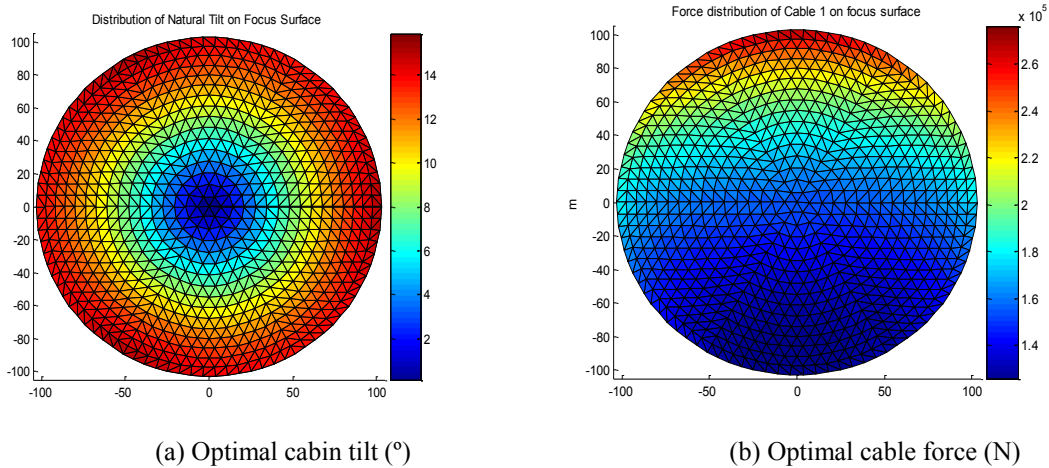

(a) Optimal cabin tilt (º)    (b) Optimal cable force (N)

Figure 12 Distribution of cabin tilt and cable force on the focal surface

### 3.3.2 Fine drive system

The fine drive system directs the receiver platform to the desired orientation and compensates the remaining positioning errors left by the main position loop. The Stewart stabilizer, also called hexapod, is the main mechanism to fulfill this task. It has to support at least a payload about 3 tons plus the structural weight of receiver platform. To meet such requirements, both its size and weight have to reach a surprisingly large level, not to mention other specifications like stiffness, stroke and power supply demanded by its 6 legs. Therefore, it is highly desirable to minimize the work space of the Stewart stabilizer. Since the main part of the work space comes from the rotation of receiver platform, it is advantageous to add two extra rotators between the first and second control loop, namely A and B axes (AB axes). The rotator can help the Stewart stabilizer to compensate the steady-state difference between natural cabin tilt and the required pose which reaches its maximum on the edge of the focal surface (Figure 12a). The Stewart stabilizer is used only for fast small deviations caused by the remaining errors left from the main position loop while the compensation of the rotator is basically quasi-static in the tracking mode.

### 3.3.3 Closed loop control

As aforementioned, the overall control system is made up of main position loop, fine drive loop and probably an additional vibration control loop as well if active damper is used to add artificial damping.

The main position controller gets the feedback information of the actual cabin position, compares it with the desired "reference" position, and then decides what corrections should be taken by the capstan drives. The controller bandwidth below the basic mechanical resonance modes of the cables (0.18 Hz) is critical to achieve the final accuracy of 10mm as discussed on control stability. Besides, a problem special to the main position controller may occur in that it has to handle the coarse compensation of cabin position as well as a part of cabin pose to satisfy the natural cabin tilt, as shown in Figure 12. However, without the information of natural cabin tilt, the angular control difference is not easy to calculate from sole position feedback. Force feedbacks of 6 cables are thus necessary. They can be used to keep cable force above a set threshold for coarse pose control and also to help get natural cabin tilt for the more accurate pose control. As both two types of adjustments are done via six capstan actions, their respective proportion in the actions is also a parameter. The position control takes up a main part in the baseline controller design of the main position loop. Accurate pose control is significantly influenced by the accuracy of force sensor and pulsating cable forces caused by vibrations. Its reliability is still under investigation.

The fine drive control system consists of two sequentially arranged loops for the rotator and the Stewart stabilizer. The two control loops have very different tasks:

1. The rotator has to compensate the angular difference between the natural cabin tilt and the required direction of the receiver platform. The correction is handled by rotations about two sequentially arranged axes with movement range about ± 24°. The corrections are needed only for sidereal tracking mode and are therefore slow.

2. The Stewart stabilizer has to correct deviations of the main position controller in the case that its function is not accurate enough, particularly in regard of dynamics. Therefore the required corrections are small, but have to be relatively fast and have to be executed in all six rigid body degrees of freedom. This is achieved by parallel arrangement of six linear screw type actuators.

### 3.3.4 Mechanical design for power supply and signal transmission

Since there is no stiff connection between the ground and the cabin, there are difficulties in supporting the power supply and transmitting signals to/from the equipments in the cabin. Due to the huge telescope size and cabin movement range, it is not practical to let the related power cables and signal cables suspended directly in the air or to span an additional supporting line between two diagonal towers. The basic concept is to use existing steel ropes as hanging-up structures. Two candidate design, curtain-style hanging-up and composite cable, are proposed to adapt the support in real time to the large cabin movement range.

Figure 13 shows a curtain-style hanging-up mechanisms connecting power/signal cables with an existing steel rope. Transverse section of Composite cable is shown in Figure 14. In 2010, curtain-style hanging-up mechanism was selected as the final design because of its simplicity and high reliability.

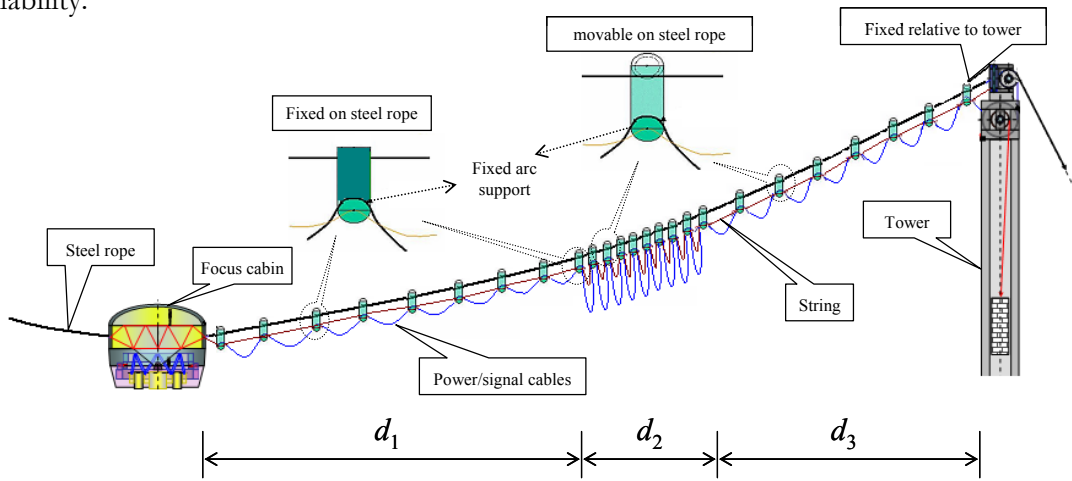

Figure 13 Curtain-like hanging-up mechanisms of power/signal cables

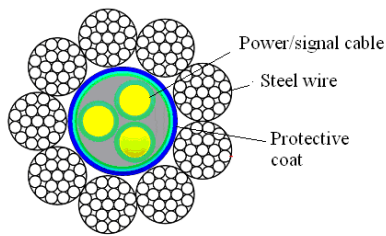

Figure 14 Sketch of transverse section of composite cable

### 3.3.5 Simulations and model experiments

Based on different design concepts, quite a few experimental models have been built and studied since 1999, including the 50m scale models in Xidian University, Tsinghua University and Miyun Station, respectively. A comprehensive simulation study was subsequently carried out in Tsinghua University to validate the 50m scale model in Miyun Station (Figure 15). Moreover, an end-to-end simulation study was finished in 2007 under Sino-German cooperation. The available engineering experience and analysis tools are combined to a 1:1 end-to-end model, which is able to describe the overall system from the commands of telescope operator up to the output from science observations.

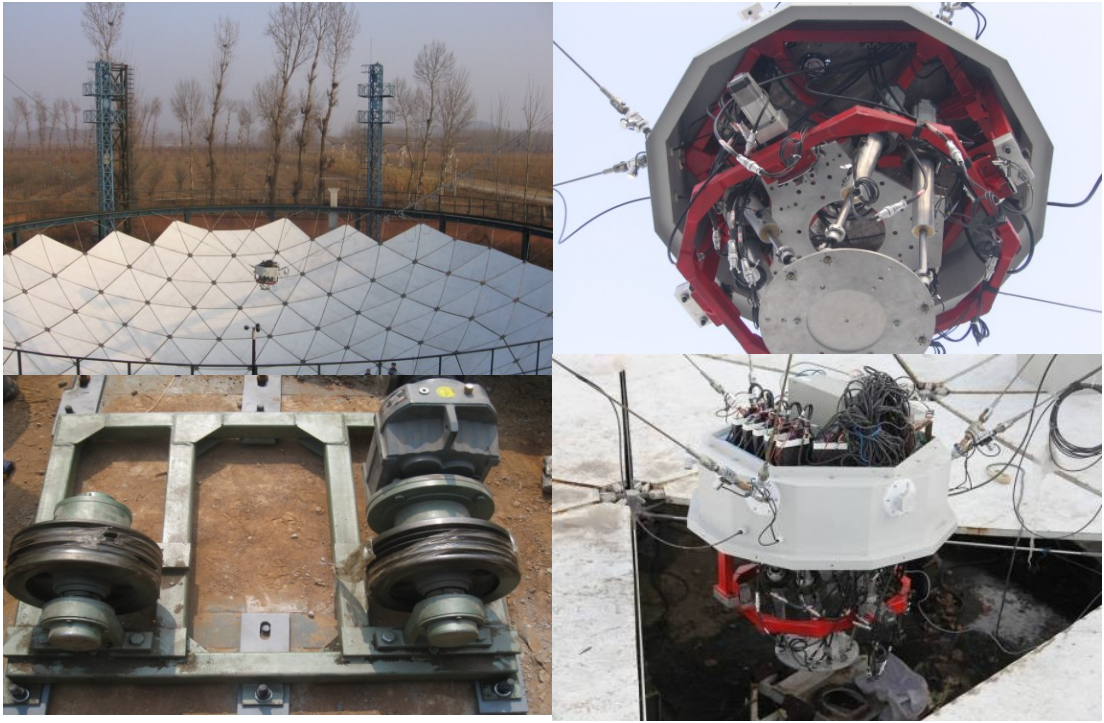

Figure 15 New demonstrator with complete mechanisms at Miyun station.

### 3.3.7. Requirement from FAST sciences

In order to track a celestial object, the feed should move with speed high enough (about 15° per hour in the limiting case) in the focal plane while keeping 8" pointing accuracy. A Stewart platform with added rotator is designed to fulfill such requirements.

### 3.4. Receivers

### 3.4.1 The Specification of Receivers

FAST can be seen as a a primary focus parabolic telescope. Unlike conventional steerable parabolic telescope, where the gain is normally a function of elevation due to gravitational deformation of mechanical structures, FAST is expected to have constant gain at different zenith angle since the panels stay at almost the same tilting angle during an observation. The illuminated aperture is of 300 m diameter and the spillover will normally be on metal reflector panels instead of hot ground. Such arrangement gives more freedom while optimizing illumination pattern of the feed, e.g., higher tapering at the edge of the illuminated parabolic aperture may give higher gain, while keeping the system temperature, more or less the same.

The layout design for FAST receivers has been drafted and being continuously revised through a long-standing cooperation agreement signed by the Jodrell Bank Observatory (JBO) and NAOC. The operational bands cover a frequency range of 70MHz – 3GHz, with the possibility of upgrading to C and X bands. In the current layout, there are 9 sets of receivers as listed in Table 2. The layout also includes scientific backends, time and frequency standard, monitoring/diagnostics of the receivers. This receiver layout provides weight and size of the receiver frontends, which is a very important input parameter for designing the feed cabin suspension system.

**3.4.2 Receiver frontends**

The multi-beam receiver at L-band is one of the most important receivers for FAST. It is mainly used for HI, pulsar and other survey observations. The phase array feed (PAF) technology may be able to generate ~100 beams, but the receiver noise temperature of the current PAF is much higher than traditional horn receivers, and beam forming network also add complexity to the receiver system. With the sensitivity being the prime concern, multi-horn receiver is chosen for FAST. One of the successful L-band multi-horn receivers in astronomy is the Parkes 13-beam receiver built by Commonwealth Scientific and Industrial Research Organisation (CSIRO). It covers a frequency range of 1.23-1.53 GHz. The CSIRO also provided a 7-beam multi-horn receiver for Arecibo telescope. Compared with Parkes, FAST has a larger f/D, which allows for a feed array with more horns since the coma effect is less severe. A tri-lateral collaboration among FAST project of NAOC, Jodrell Bank Centre for Astrophysics (JBCA) of Manchester University and CSIRO has been established for a designing study of a 19-beam multi-horn receiver. The frequency range for the 19-beam receiver listed in Table 2 is the same as the Parkes 13-beam receiver. One of the purposes of the trilateral design study is to investigate the feasibility of a wider frequency coverage. In 2009, the study has demonstrated the feasibility of 19-feed array which covers frequency range of 1.05-1.45GHz. The study also included the orthomode transducer (OMT). The three parties are now considering prototyping and detailed design for this 19-beam multi-horn receiver.

A single pixel receiver at L-band with much wider bandwidth is planned for pulsar timing, international VLBI and the other observations of targeted objects. Corrugated horn and quad-ridge OMT may be used. A high temperature superconductor (HTS) hybrid may be used to convert linear polarization from the OMT to circular polarization output. In S band, a single pixel receiver is planned. Scalar horn and quad-ridge OMT will be used. Again, a HTS hybrid is used for extracting circular polarization.

Low frequency range is divided into 4 octave bands and their feed design will adopt those existing technology used by telescopes such as ATA, LOFAR and WSRT. Fewer sets of feeds with wider frequency coverage have also been considered, e.g. a scaled version of the eleven feed designed for VLBI 2010. Cryogenics is not considered at frequencies below ~500MHz, where the system noise temperature is dominated by sky background. Two narrow band receivers, set 5 and set 6 in the table 2, are designed for telescope adjustment and early sciences.

The frontend of receivers at frequency higher than ~500MHz will be cooled by using Gifford-McMahon cooler. The compressor will be mounted on the first stage platform of the feed cabin. The compressor may be hung on a gimbal-like mechanism to keep them horizontal. The refrigerator attached to the associated receiver will be mounted on the stabilized platform. The helium will be supplied via a mixture of stainless steel flexible and rigid tubing between compressor and refrigerator.

### 3.4.3 IF transmission using optical fiber

Optical fibers will be used to transmit the IF signal to the ground level. Wideband analogue link on optical fiber is considered in the current layout. Multiplexing may be adopted to reduce the number of fibers required (Figure 17).

### 3.4.4 Receiver arrangement on the stabilized platform

The frontend of the receivers will be mounted on the stabilized platform (Figure 16). Provision has been made for mounting the L-band multi-horn receiver at the center of the platform. All the other receivers are mounted surrounding the multibeam receiver. Several spare mounting points are planned for future upgrades. The size and weight of a PAF receiver at L-band is also considered when determining the payload of the stabilized platform. The LO/IF electronics and optical fiber modulation system will be mounted on the first stage platform of the feed cabin.

### 3.4.5 Scientific backends

Digital backends will be adopted for various scientific purposes. The backends will be put in a screened room in order not to interfere the observation. High speed A/D converter and FPGA based boards developed by Center for Astronomy Signal Processing and Electronics Research (CASPER) group will be used for spectral line, pulsar and search for extra-terrestrial intelligence (SETI) observations. Digital base band converter (BBC) and disk based MK5 will be used for VLBI terminal. A general purpose backend based on PC-cluster is also considered for future development.

Table 2　9 sets of receivers

| No | Band (GHz) | Number of Beams | Polarization | Cryo $T_{sys}$(K) | Science |
|---|---|---|---|---|---|
| 1 | 0.07 – 0.14 | 1 | RCP LCP | no 1000 | High-z HI(EoR), PSR, VLBI, Lines |
| 2 | 0.14 – 0.28 | 1 | RCP LCP | no 400 | High-z HI(EoR), PSR, VLBI, Lines |
| 3 | 0.28 – 0.56 | 1 or multi | RCP LCP | no | High-z HI(EoR), PSR, VLBI, Lines, |

|   |             |                          |         | 150 | Space weather, Low frequency DSN |
|---|-------------|--------------------------|---------|-----|----------------------------------|
| 4 | 0.56 – 1.02 | 1 or multi               | RCP LCP | yes<br>60 | High-z HI(EoR), PSR, VLBI, Lines, Exo-planet science |
| 5 | 0.320 – 0.334 | 1                      | RCP LCP | no<br>200 | HI, PSR, VLBI,<br>Early sciences |
| 6 | 0.55 – 0.64 | 1                        | RCP LCP | yes<br>60 | HI, PSR, VLBI,<br>Early Sciences |
| 7 | 1.15 – 1.72 | 1 L wide                 | RCP LCP | yes<br>25 | HI, PSR, VLBI, SETI, Lines |
| 8 | 1.23 – 1.53 | 19 L narrow multi-beam   | RCP LCP | yes<br>25 | HI and PSR survey, Transients |
| 9 | 2.00 – 3.00 | 1                        | RCP/LCP | yes<br>25 | PTA, DSN, VLBI, SETI |

### 3.4.6 Requirement from FAST sciences

Searching for low frequency spectral lines by using FAST requires a full coverage of the spectra. The receivers are designed to cover the 70MHz~3GHz band continuously, in which L-band 19-beam receiver is the main instrument, required by Milky Way and nearby galaxy HI studies and pulsar surveys.

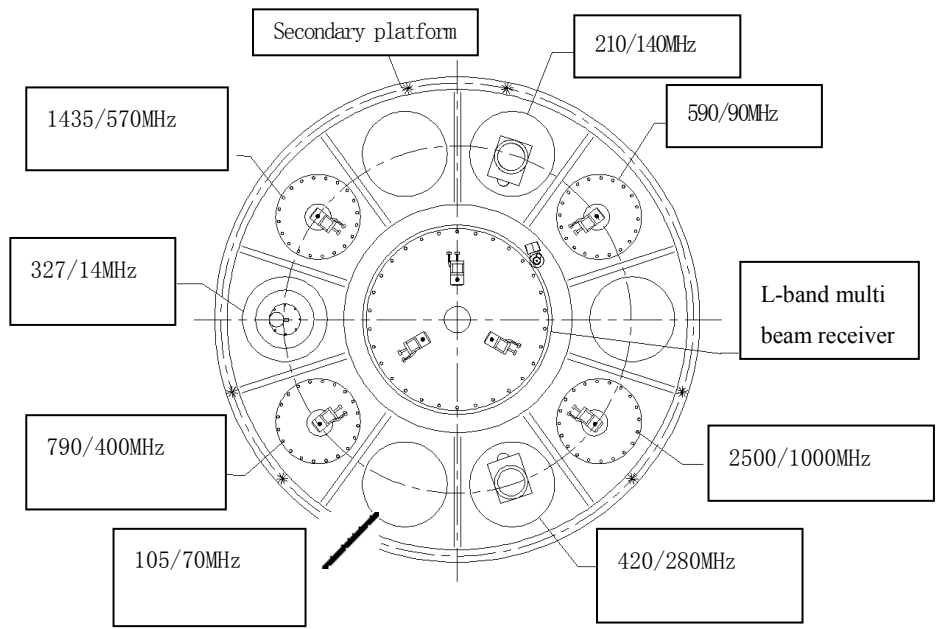

Figure 16 Receiver arrangement on the secondary stabilized platform. The center frequency and bandwidth for each receiver are indicated in the text boxes. Two mounting points (the empty circles) are planned for future upgrading.

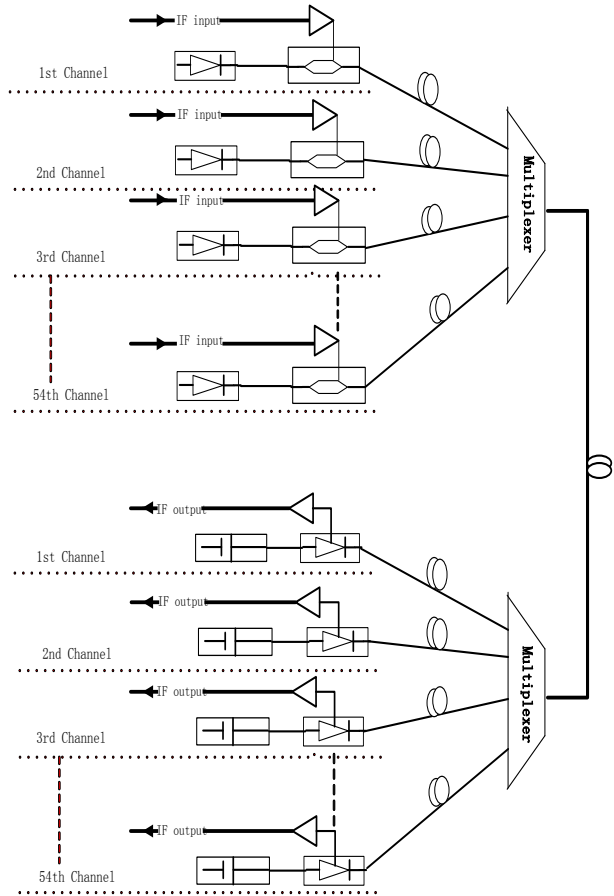

Figure 17 Optical fiber link used to transmit the IF signal to the ground level observing room. The E/O part (upper) and O/E part (lower) are shown.

## 3.5 Measurement and Control

With ample adjustable components, the measurement and control subsystem is essential for FAST operation.

### 3.5.1 System structure and indicator

Measurement and control system consists of four parts: overall control system, benchmark measurement, measurement and control of feed cabin suspension, and measurement and control of reflectors. Requirements of specific indicators are listed as follows:

    1. Position benchmark: benchmark points' relative position with an accuracy of 1mm;
    2. Time benchmark: 1ms;
    3. Position measurement of base points: 2cm;
    4. Control precision of feed cabin suspension: 10mm;
    5. Control precision of the main reflector node: 2mm.

### 3.5.2 Benchmark measurement

There is no rigid connection between the main reflector and feed support of FAST. The coordinated movement needs high-precision time and position benchmark. Time benchmark require 1ms measurement precision. GPS granting time, computer keeping time and the entire network checking time are designed to satisfy the synchronization requirements of FAST's measurement and control. The aggregated measurement time precision is much better than 1ms. According to position benchmark, benchmark points' relative position precision must reach 1mm. The layout design of the benchmark network and points has been completed (Figure 18). Primary control network has been built.

### 3.5.3 Measurement and control of feed cabin suspension

There are towers of about 100 m height supporting 6 cables which drive the feed cabin to move on the focal surface. The AB axes and Stewart platform fine-adjusting mechanical structure are placed in feed cabin, which improve the positioning accuracy of platform carrying the receiving system to 10mm. Large-scale, high precision and high sampling rate measurement equipments are essential. Feed cabin compartment include three parts: star frame, AB axes and fine-adjusting system. Star frame is dragged by cables to complete coarse position determination. AB axes mechanical structure completes the coarse attitude determination. Fine-adjusting system completes the precise adjustment of the position and attitude.

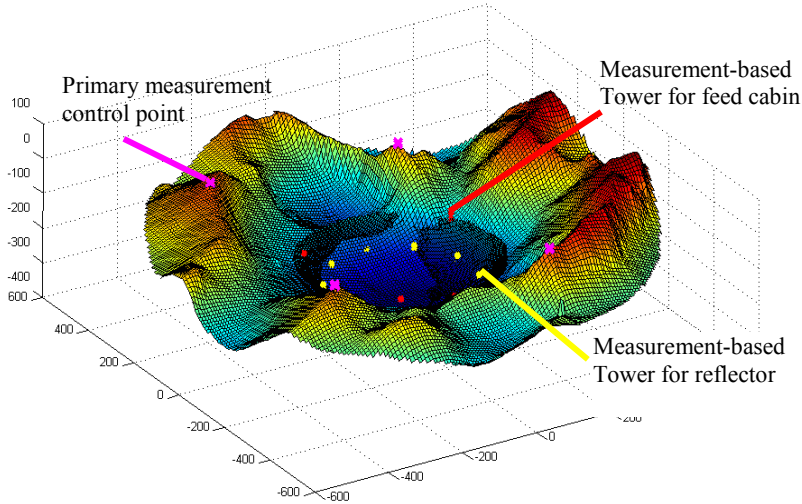

Figure 18 The layout design of the benchmark network and points.

### 3.5.3.1 Feed support system measurement

Control system should achieve closed-loop control. Based on the designing principle of loose coupling and results from engineering testing, coarse-adjusting system can be independent from fine-adjusting system completely.

The position determination precision of the coarse adjusting measurement is 100mm. The attitude determination precision in 10 meters scale is 33' (0.5°). Laser tracking and GPS measurement are employed (Figure 19).

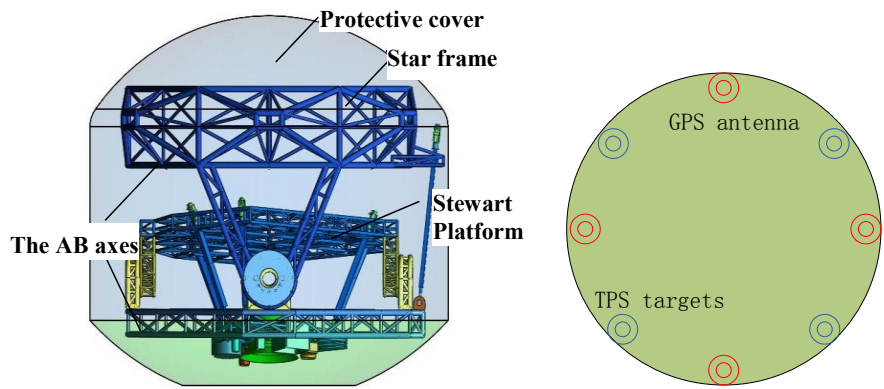

Figure 19 Feed cabin structure and targets in the star frame

Fine-adjusting mechanical structure Stewart platform is composed by upper platform, lower platform and 6 branched-chains. The position precision of the below platform is 10mm and that of the fine-adjusting measurement system and the distributing measurement is 5mm. Precision laser measurement technology, precision inertial measurement technology and precision photography measurement technology are the alternative methods for fine-adjusting measurements. The following figure is Stewart platform structure and distribution map and 12 mounting interfaces in the preliminary design (Figure 20).

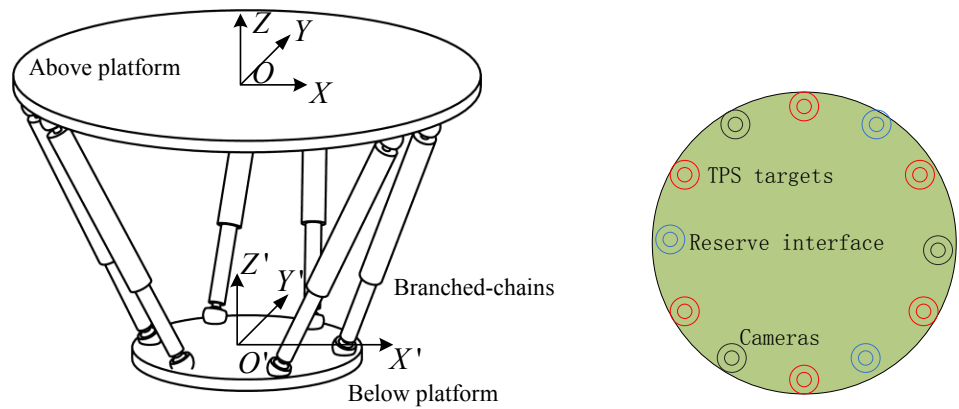

Figure 20 Stewart platform structure and targets in below platform

### 3.5.4 Measurement and control of the reflector

The surface precision of 500-meter diameter reflector is required to reach 5mm. There are 2300 cable network nodes on the reflector. The maximum tracking speed of the nodes is 0.7mm/s. Range of motion is 95cm. Measuring more than 1000 nodes needs be done within 1 minute. Measurement precision needs to reach 2mm.

Digital close range photogrammetry is adopted to accomplish the real-time measurement of the main reflector. Figure 21 gives a sketch of the stereo vision measurement system. It's not necessary to scan the entire reflector all the time during observations, only the measurement of 1000 notes located in the corresponding region of the radio source needs to be started.

The layout of the three-dimensional digital measuring equipment on the edge of the top of the reflector is shown in Figure 22. As the three-dimensional digital measuring equipment has a two-dimensional turntable, the measurement programs have great flexibility. The simulation results shows that all the equipments can observe the same region at the same time, and the measurement of the entire 1000-node region can be finished within 1or 2 minutes.

The targets are mainly divided into active and passive. Different types of targets have been used in the experiment. The targets are shown in Figure 23.

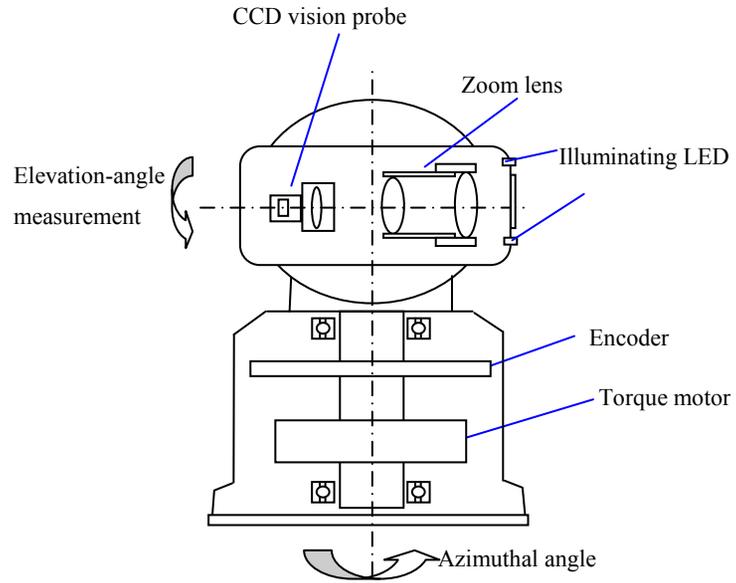

Figure 21 The sketch of Super-high-accuracy stereo vision measurement system

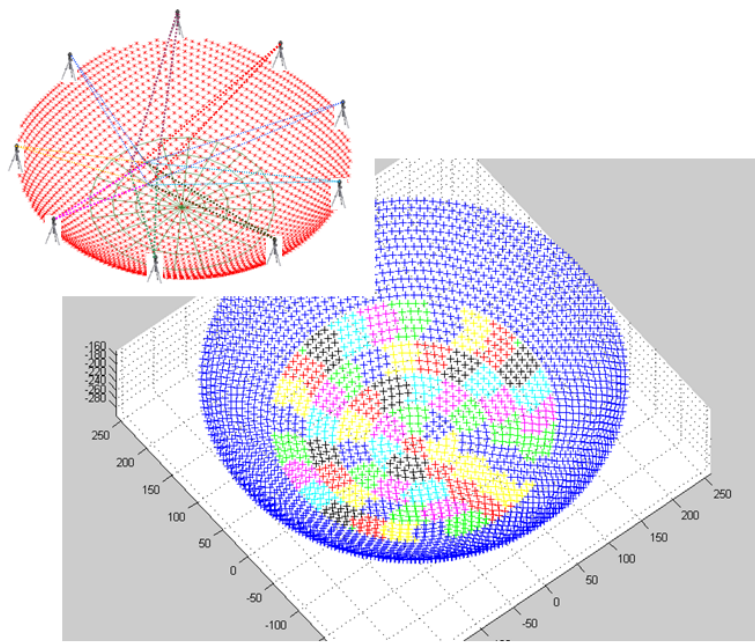

Figure 22 The layout of digital measuring equipment on the edge of the top of the reflector

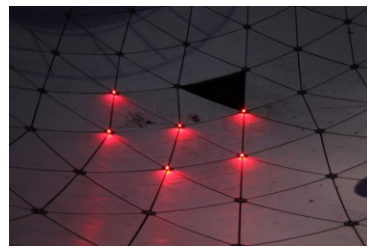

a). Active target

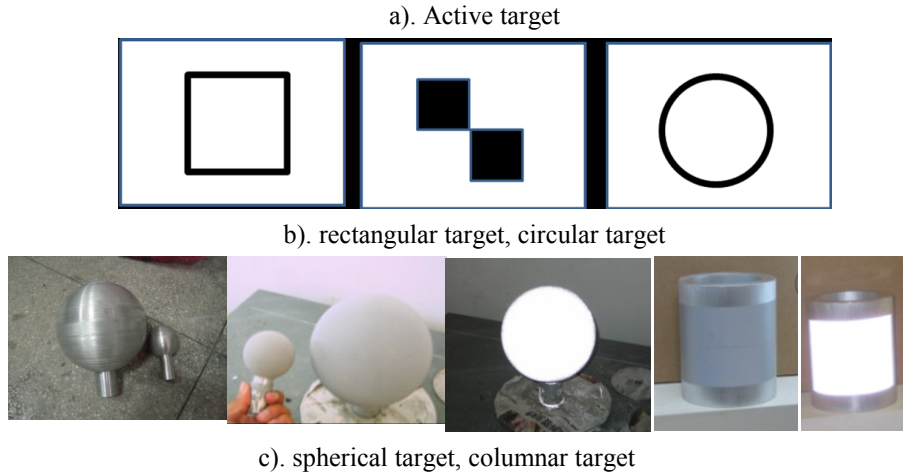

b). rectangular target, circular target

c). spherical target, columnar target

Figure 23 Targets

The control system of the reflector is a three-layer structure. The top layer is the central control unit. The middle layer is the exchanging layer, and the bottom layer is the actuator controller. The controller receives and processes the feedback node location from the measuring system and the directive and control data from the collectivity controller to executive the motion control of the actuator. System constitution diagram is shown in Figure 24.

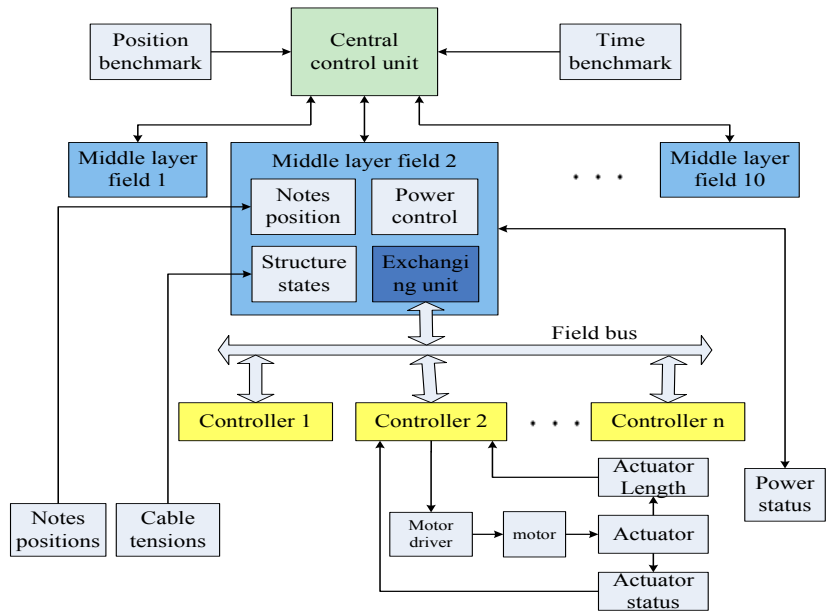

Figure 24 System Constitution Diagram

### 3.5.5 Requirement from FAST sciences

The feed cabin and main reflector must operate in phase without rigid connecting structure to realize pointing and tracking. Measurement and control system is essential to reach this goal. It also provides the status of the telescope, which is crucial to calibration and data preparation.

## 4. FAST Sciences

Being the world largest filled-aperture telescope located at an extremely radio quiet site, FAST will have extraordinary science impact on astronomy. It will have potentials to revolutionize other areas of the natural sciences. Compared with its closest precursor—Arecibo, FAST has a factor of 3 better raw sensitivity and a factor of 10 higher surveying speed. FAST will also cover two to three times more sky area thanks to its innovative design of active primary surface. A science instrument with an order of magnitude improvement in any capacity and being able to explore new parameter space will likely generate unexpected discoveries. Our planning, however, is limited by what we know from the existing instruments. Thus, this section contains largely logical extension of what we are doing now with single dish radio telescopes. We mention some projects of more speculative nature toward the end.

We organize our discussion as following. First, we lay out the fundamental scientific questions relevant to FAST (4.1). Second, we discuss in more detail three categories of objects, the Milky Way ISM (4.2), HI in external galaxies (4.3), and pulsars (4.4), which will be the main targets of FAST observations. Last, we briefly discuss some specific types of projects and/or operation modes (VLBI 4.5 and SETI 4.6), to which FAST is capable of making unique contributions.

### 4.1 Fundamental Questions for a Large Single Dish Radio Telescope

The origin of the observable universe, the origin of our world with the Sun and the Earth, and the origin of intelligent life are overarching questions of natural sciences. FAST, with its unparalleled collecting area, state of art receiver systems, and the digital backend of which the technology development largely follows the Moore's law, has a unique window for contribution through precise measurements of matter and energy in the low frequency radio bands.

At radio frequencies, a large single dish telescope is capable of observing the main component of cosmic gas, atomic hydrogen (HI), from local universe to moderate redshifts. The gaseous galaxies can be either bright or totally dark in optical bands depending on their history of star formation. Therefore, a complete census of gaseous universe through blind surveys provides information of cosmology and galaxy evolution independent of those based on star light. One exciting development of research in cosmology is the apparent success of ΛCDM simulations (e.g. [9]) based on models in producing large-scale structures of dark matters. This is accomplished without knowing the actual content of neither dark matter nor dark energy. The critical test of such models and associated cosmology is to compare predicted structures to observable matters. One current mystery is the so called "missing satellite problem", i.e., the lack of detection of low mass halos predicted by dark matter simulations. Given the uncertainties in our knowledge of star formation (discussed later) and the very rudimentary treatment of star formation in simulations, the stellar content of these halos is essentially unknown. Therefore, by providing a census of HI complete to relatively low mass limit along with other large radio telescopes, FAST can significantly improve our knowledge of the origin of the universe.

The formation of stars and planets is a fundamental process through which our world was made. The Sun is generally categorized as one of the low mass stars which comprise the majority of stellar masses in the Milky Way. The long lifetime and large number count of low mass stars provide the stability and possibly the time scale for the evolution of our world. Young stars are usually directly seen in the infrared to UV bands, radio telescopes on the other hand are particularly good for studying the conditions of ISM, and the birthplace of stars. For Galactic observations, the collecting area and L band instruments of FAST, are capable of capturing HI gas in a band of over 1000 km/s wide and with velocity resolution better than 0.1 km/s ($\Delta\nu \sim 0.5$ kHz). These unique capabilities in the spectral domain of large radio telescopes enable detailed studies of low mass star formation, especially in nearby regions. One primary area is to study the cold population of atomic hydrogen through absorption signatures, HI Self-Absorption (HISA; e.g. [10]), HI Narrow Self-Absorption (HINSA; e.g. [11]), and absorption against background continuum sources. The vast spectral dynamic range facilitates the separation of cold gas from the general galactic HI background in the velocity space and provides information on their excitation conditions. These absorption features trace histories of the cooling of atomic hydrogen and the combination of hydrogen into molecular form, which are prerequisite steps of star formation. High mass stars, although occupying minority portions of mass and having a much shorter lifetime, dominate the dynamical processes and evolution of matter on cosmological and galactic scale. The nuclear synthesis and dust formation associated with massive stars account for the majority of elements heavier than Helium that can be seen today. The spectral dynamic range of four decades provided by FAST enable detailed looks into the dynamical processes and structures produced by massive stars. The so-called super bubbles and chimneys are examples of dynamically coherent structures likely associated with both the formation and destruction of massive stars. Accounting the matter and energy content of these structures provide a global view of the evolution of ISM propelled by star formation. Radio telescopes with its precision in frequency and time measurements are also unique for capturing pulsars, one of the end products of massive stellar evolution. Arecibo and other current radio telescopes are dedicating significant share of their time for pulsar searches. FAST would extend these searches to unprecedented sensitivity and sky coverage, thus provide new accounting of massive stars throughout their history in the Milky Way (e.g. probability of gaining escaping velocity). Pulsars are also being treated as laboratories of physics under extreme conditions in terms of density, magnetic field, and for strange states of matters. The precise nature of timing signals provided by pulsars could also be used as a cosmic probe of gravitational waves, potentially opening a new observing window to the universe.

Discoveries of exoplanets have generated tremendous interests and rekindled the discussion of whether intelligent life is unique to Earth. Most of the $\sim$ 500 known exoplanets so far are detected through indirect measurement of stellar light in either radial velocity or transiting light curves. In the Solar system, Jupiter generates non-thermal emission below 50 MHz where the Sun is relatively radio quiet. The sensitivity of FAST will be enabling to make direct observation of exoplanets in meter wave band. If detected, such radio emission opens a new window through which we can study the magnetic field and the physical environment of alien worlds. SETI will be another important component in FAST's efforts to understand the origin of intelligent life. The key aspect of radio SETI is a search of narrow band radio signals toward nearby stars. The unparalleled sensitivity and sky coverage of FAST will make possible a SETI survey that significantly superseding the volume of existing surveys.

## 4.2  The Milky Way ISM

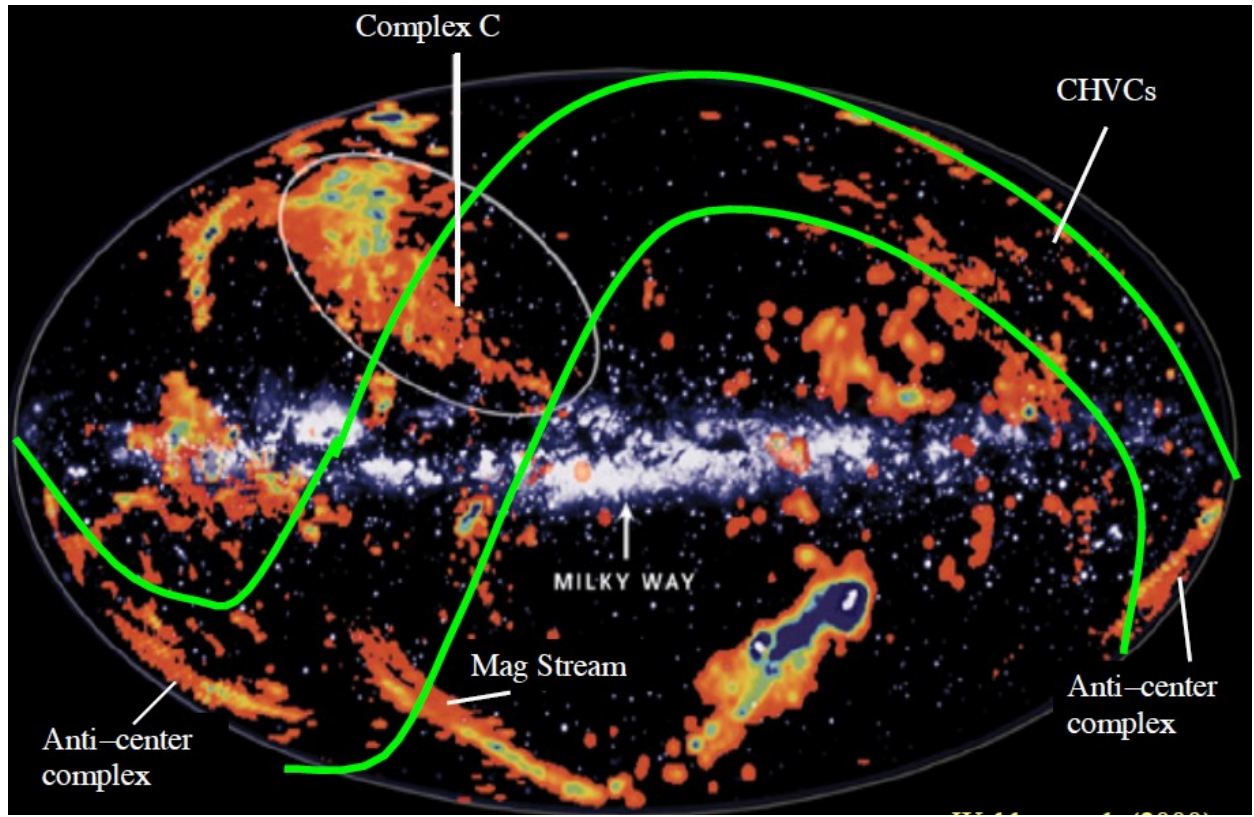

Figure 25 Distribution of the halo clouds with green lines indicate Arecibo sky coverage, which will be enlarged by a factor ~ 2 by FAST. This figure is adopted from the white paper "Galactic Astronomy with the Arecibo L-band Feed Array" [12]).

FAST will be a powerful instrument in studying the Milky Way, and in particular the ISM, through observing different kind of radio signatures including the 21 cm line, radio continuum, and recombination lines.

The hyperfine transition of atomic hydrogen at 1420.405 MHz (21 cm) is the main probe of the most abundant element of matter in the universe. Systematic surveys of the 21cm line, both in the Milky Way and external galaxies, are one of the main science drivers of FAST.

The all sky Galactic HI survey currently available is at a resolution of ~30' [13]. For Galactic disk, all major radio interferometers around the world have been carrying out coordinated surveys, including Canadian Galactic Plane Survey (CGPS, [14]), VLA Galactic Plane Survey (VGPS, [15]), and Southern Galactic Plane Survey (SGPS, [16]). These surveys cover a large fraction of longitude of the disk within latitude of ~ +/- 1.5 degrees. The angular resolution of these surveys is about 1" and the velocity resolution is about 1 km/s. A consortium of radio astronomers have been using the Arecibo Focal Plane Array (ALFA) to make large HI maps toward selected regions of the Galaxy, including the Taurus molecular cloud and the Magellanic stream. The angular resolution of Arecibo HI observations is about 3.5' and velocity resolution of about 0.1 km/s. With the 19-beam L-band focal plane array, FAST will be about 5 times faster than Arecibo in mapping Galactic HI with ~30% better angular resolution and can observe 2 to 3 times more areas of the sky than Arecibo. The velocity resolution of FAST will be comparable to Arecibo, and significantly better than that achieved by interferometers. Therefore, FAST HI surveys will provide a better census of the neutral ISM than existing surveys combined. The angular and spectral resolution of FAST is comparable to those of large-scale surveys of CO, the major tracer of molecular gas in the Galaxy. FAST surveys thus enable direct comparison of CO and HI and subsequent studies of the evolution of ISM from atomic to molecular phase. Such comparison has not been possible due to the limited declination range of Arecibo, especially in some well-known star forming regions, including Orion and others further south. A sensitive census of atomic hydrogen of the Galaxy with broad dynamic range is also necessary for studying the content and origin of the high velocity clouds, Galactic halo and its connection to the disks. In short, a new inventory of Galactic HI gas from FAST and its comparison with molecular gas (where available) will provide important observational evidences for studying Galactic structures and evolution of ISM.

Radio continuum is another fundamental target which FAST should be able to observe on a Galactic scale. The simultaneous bandwidth of ~ 500 MHz of the L-band focal plane array would be sufficient for making continuum surveys of a large fraction of the northern sky. A particular interesting topic is the polarized structures in radio continuum. Low frequency Galactic radio continuum is well known to be originated from Galactic magnetic field through synchrotron emission. At higher angular resolution, e.g., in L band from Effelsberg, complex structures have been found in polarized intensity, which are not correlated with total intensity or consistent with the general picture of a smooth Galactic B field. The detected polarization seems to be dominated by structures and effects such as Faraday Screen of the medium through which the synchrotron emission travels. One of the largest published polarization surveys in L band has been carried out using the 25.6 m telescope at the Dominion Radio Astrophysical Observatory [17]. The Galactic ALFA consortium also has a key program to map a substantial fraction of the Arecibo sky. FAST, given its advantage in mapping speed and sensitivity, will make significant advances in imaging the polarized radio sky. Better knowledge of polarized radio continuum and better understanding their origin will not only help us in studying ISM, but also in provide better foreground template for studying polarized cosmological background emissions. Large continuum surveys, when done in multiple overlapping passes, will also be instrumental in discovering transient objects. These sources represent likely locations of dynamical events, which could provide probe to fundamental physics.

FAST has continuous spectral coverage from 70 MHz to 3 GHz. Other than the HI line and radio continuum discussed in some details above, there are also other diagnostics of ISM, such as recombination lines, masers, and molecular lines, of which FAST will obtain significant data. About 20 percent of the molecular lines are observed in centimeter and decimeter bands. The operating frequencies of FAST cover lines from 17 kinds of molecules, including hydroxyl (OH), methyl alcohol ($CH_3OH$), formaldehyde ($H_2CO$), etc. FAST is able to make deep surveys of the molecular masers in our galaxy as well as in the other distant galaxies. Most distant OH mega-masers have been observed by the Green Bank Telescope at z ~ 0.765[18]. A similarly bright gravitationally lensed object would be detected by the FAST at z > 1. We could also expect to discover some long carbon chain molecules in Orion, the richest spectral source, which is not in Arecibo's sky.

The combination of these studies of the Milky Way could provide a comprehensive picture of the content, dynamics, and physics of Galactic ISM.

### 4.3  HI in External Galaxies and Cosmology

FAST will be apt to probe the faint end of the HI mass function in the very local universe. Compared with existing projects carried out by ALFA, it will make one order of magnitude improvement in terms of surveying speed. These shallow surveys will enable us to investigate the extent of HI disk, answering the question of the truncation of outer HI galactic disk; to extend rotation curve to unprecedented distance, mapping out the distribution of dark matter in the local group; and to detect Cold Dark Matter Satellites and possible HI companions [19]. FAST surveys can also help determine the populations of the High Velocity Clouds (HVC) around other galaxies that is similar to our Galaxy, but for which the distance is not ambiguous. Another important subject would be looking for the low brightness galaxies in the local void to see whether there are many such objects in regions of low cosmic density.

Astronomers have long relied on optical photons to detect the matter content in the universe. To directly test the validity of ΛCDM simulations, structures made in dark matter have to be compared with observed matter, mostly based on galaxies in star light. Such comparison, however, reveals discrepancies, one of which is the so called "missing satellite problem". The number of observed low mass haloes are much less than dark matter halos predicted by simulations. The resolution of this problem and further investigation of cosmology requires a more complete inventory of matter than just the star light.

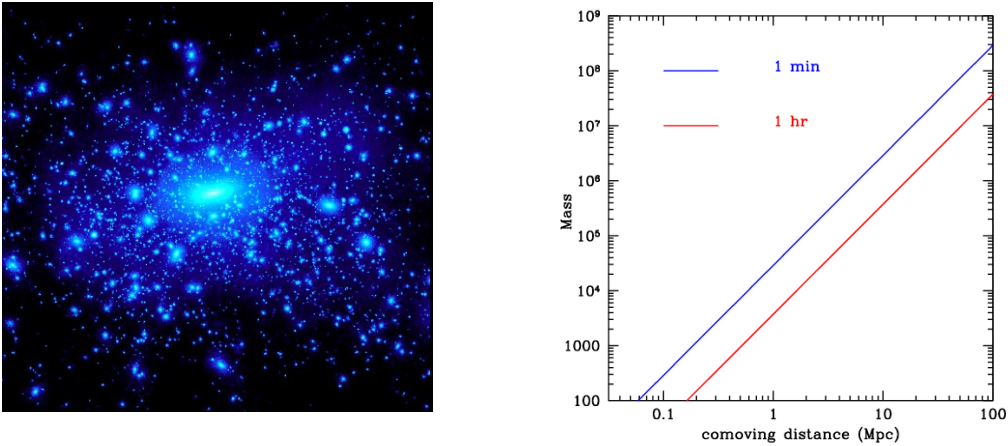

Figure 26 HI detection with FAST

The high sensitivity and multi-beam enable FAST to slice the matter power spectrum by surveying large number of galaxies in a large volume. With an interference-free observing period of 1 h, FAST can detect 'gaseous' galaxies (with $M_{HI} \sim 10^{10} M_\odot$) in HI line out to a redshift of 0.7, which can constrain the equation of state of the Dark Energy and help to understand the evolution of galaxies. In blind survey mode and 6000 second integration per field, FAST expects to detect ~42,000 galaxies with $<z> \sim 0.3$ in 30 days [2]. Figure 26 illustrates the HI detecting power in terms of mass (Dr X. L. Chen, private communication). The left panel shows the dark matter distribution in a galaxy with a size like our galaxy simulated by Kravtsov [20]; the right shows the detectable mass of a galaxy at different distances using FAST with one minute and 1 hour integration time.

Although difficult to detect a single galaxy in HI at higher redshift, say z~1, it is possible to detect the HI emission from an assemble of galaxies [21]. The HI power vs. redshift will be valuable information for studying galaxy evolution. Another survey mode is to target distant sources with known redshift. Catinella et al. 2008 [22] have demonstrated detectability of massive disk galaxies at Z~0.25 with a few hours Arecibo integration. FAST will improve such targeted surveys by one order of magnitude in terms of time requirement and/or sample size.

### 4.4 Pulsars

The high sensitivity and larger sky coverage compared with Arecibo make FAST a powerful tool for detecting pulsars at large distances, such as millisecond pulsars, binary pulsars, double pulsars, extragalactic pulsars, etc. It is estimated that FAST equipped with multi-beam receivers would detect thousands of pulsars in the Milky Way Galaxy in less than a year of observing time. In such a new large scale survey, moreover, extremely interesting and unknown exotic objects may yet wait for discovery by the sensitive FAST as the telescope is put into operation. Among these discoveries, the most exciting one should be a pulsar-black hole binary, which will provide precise information of black hole. Besides this, FAST may also found sub-millisecond pulsars and pulsars that that has mass deviating from 1.4 solar mass. This will give insight to the equation of state at supra-nuclear density, and further the strong interaction. In this way, pulsar is a unique laboratory for studying two of four kinds of fundamental forces: gravitation and strong interaction.

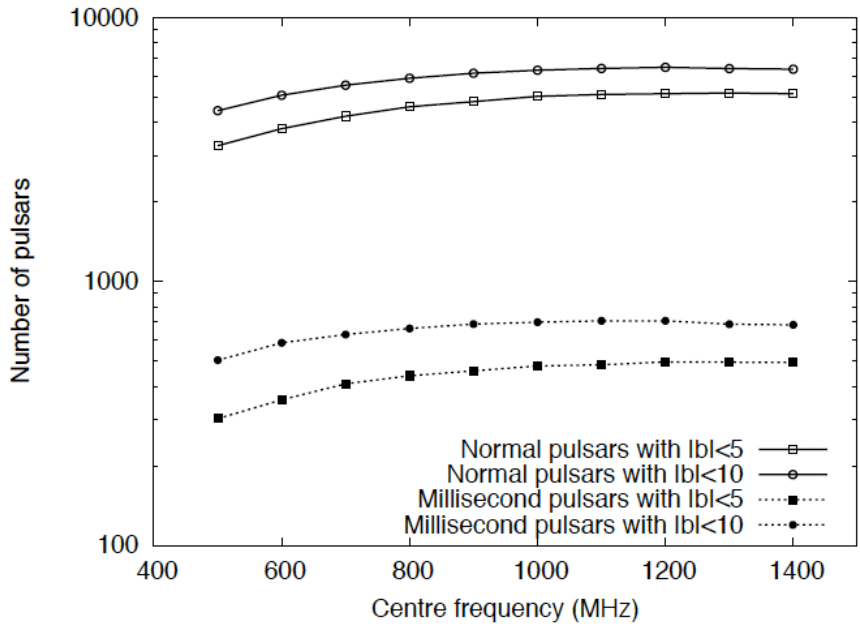

Figure 27 Number of FAST detections based on simulations by Smits et al.[23]. The simulated surveys are limited to Galactic longitude between 20 and 90 degrees.

Large number of pulsars can also be used to study the late stages of stellar evolution that lead to supernovae and compact objects, and to make sharp maps of ISM. At the early stage, even during the telescope adjustment, FAST would be able to give us opportunity in detecting some tens of pulsars in M31 with about 100 hours of observing time. Other galaxies in the local group, like M33, might also have positive detections. Using the PSRPOP package, Smits et al. ([23]) perform Monte-Carlo simulations of pulsar surveys to be carried out by FAST L band focal plane array. As shown in Figure 27, FAST is expected to detect in the Galactic plane about 5000 pulsars, out of which ~4000 would be new discoveries. The respective numbers for millisecond pulsar are about one tenth of each.

Other than the physics of compact objects, pulsars' natural precision in timing signals also provide a unique probe of the gravitational wave in space-time predicted by general relativity. FAST is capable of providing the most precise observations of pulsar timing signals, therefore, may largely increase the sensitivity and widen the spectral window for detection of gravitational waves from massive black hole pairs or the Big Bang. Figure 28 shows the future FAST limit on background gravitational wave strength. After collecting precise timing data (~30ns) for 5 years, FAST will detect background gravitational wave or put stringent constrain on current massive binary black hole and cosmic string models.

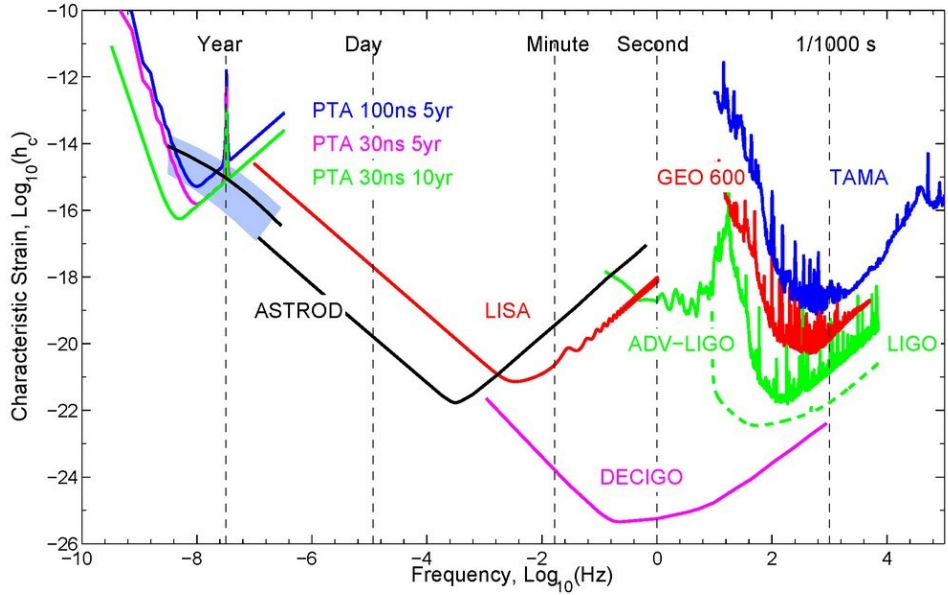

Figure 28 Future FAST (lines on top-left) limits on background gravitational wave strength. After 10 years precision timing of 20 pulsars, FAST will reach a characteristic strain of $10^{-16}$, either detect gravitational wave or put stringent constrain on current models. (Kejia Lee, private communication)

## 4.5 Very Long Baseline Interferometry

The image sensitivity of VLBA alone in L-band is 80 μJy/beam. High Sensitivity Array (HSA) which consists of full VLBA, Effelsberg and GBT 100, and VLA, gives a sensitivity of 5.5 μJy. The sensitivity would be 3.1μJy, if Arecibo is replaced by FAST. With a spatial resolution much finer than an AU-scale in nearby objects and the highest sensitivity ever achieved, this system may be able to resolve the fine structures of weak thermal sources, to get a close up of the origin and evolution of stellar sources, and even to directly image the radio loud extrasolar systems. These will promise a new avenue for scientific discovery.

## 4.6 SETI

The only practical way to contact distant civilizations may also be through radio waves. Most SETI searches concentrate on microwaves at 1-60GHz, using "free space" in the microwave window, especially the narrow band between the hydroxyl (18 cm) and neutral hydrogen lines (21 cm), called the water hole because the combination of OH and H yields water. Among international SETI teams, Project Phoenix hosted by the SETI Institute seems to be the most compelling one at radio wavelengths. Phoenix started surveying some 1000 Sun-like stars out to a maximum distance of about 100 ly since the 1990s, using the largest radio telescopes available. By increasing the target number to 5000, FAST may bring us a great opportunity.

## 4.7 Early Science Opportunities

The complexity and the innovative nature of the FAST systems pose challenges for science operation, especially in its commissioning phase. We also consider this period (expected to be ~ 6 months) an opportunity for a few dedicated "early science" projects, which aim to utilize the sensitivity of FAST before the complete suite of receivers become available and all normal observing modes can be fully implemented.

Advances in radio astronomy have enabled us forming a basic understanding of composition of ISM. Key advances include the discovery of the first known interstellar gas in HI and the first known interstellar molecule in OH. A full inventory of ISM spectral lines is of fundamental significance to this field. Line survey has been a major mode of observation in millimeter bands, given the large number of lines arising from molecules, particularly from rotational transitions. In the low frequency radio bands, the spectra survey program has not been fully explored. A recent Arecibo survey of Arp220 [24], a starburst galaxy, has been fruitful in detecting pre-biotic molecule, such as Methanamine ($CH_2NH$). A primary galactic target of spectral line survey, Orion molecular cloud, is out of the sky coverage of Arecibo. The center of the Orion molecular cloud has been a rich target for studying the content of ISM. As the nearest massive star forming region, Orion possesses both very high column density of gas and effective sources of excitation. The energy output from young massive stars creates conditions for both photochemistry and reactions enabled by shocked gas. In the early science phase, we plan to observe Orion with any available band and integrated down to mK RMS in a time scale of hours. It will be interesting to compare FAST observations of Orion to Arecibo spectra of Arp 220, and thus understand the origin of interstellar lines in both galactic star forming regions and in luminous infrared galaxies. This exploratory effort also holds the potential of detecting new molecules such as heavy negative ions e.g. $C_{10}H^-$, the lighter forms of which have only recently been discovered in ISM [25].

One major early science area is HI galaxy survey. Search for HI galaxies can be done fruitfully in fast scan mode, as particularly demonstrated by the Arecibo Legacy Fast ALFA Survey (ALFALFA, [26]). To maximize the science output in terms of detections, the survey projects favors shorter integration time as the sky coverage increases linearly with speed, but the noise decreases only in square root of time. Therefore, a large HI blind survey should only be done with the focal plane array and with the faster scan modes well implemented. In the early science period, however, we should plan targeted deep observations of a few targets. The aim of such deep integration is to measure the gas content of clusters and starburst galaxies. The HI gas in these objects is a fundamental quantity for understanding galaxy evolution and cosmology, but is poorly known due to sensitivity limits of currently available instruments. Preparation is still needed to compile a feasible target list, particularly those out of Arecibo sky and of unique characteristics. At the distance of Virgo cluster, an integration of a few hours should bring the HI detection limit under $10^6$ solar masses, which would provide an unprecedented look at the low mass end of the HI mass function.

As with HI, large scale surveys of pulsar will be done more profitably with focal plane array. The monitoring of millisecond pulsar (MSP), particularly rare, fast spinning ones should start in the early science phase. The fast spinning pulsar, with pulse period as short as ~1 ms is a unique laboratory for extreme state of matter. The sensitivity and sky coverage of FAST would enable timing monitoring of a large fraction of the known MSP. The fluctuation in timing signal, the detection of which requires a time baseline to be started in the early science phase, could provide clues to gravitation wave signal. Targeted search of pulsars in nearby galaxies is another important discovery project to be attempted as soon as possible due to the potential of detecting the first extragalactic pulsar.

In short, we expect all major science areas to be touched upon during the early science period. These projects will demonstrate the working of receiver bands, forming of primary surface, and basic observing modes. Important science results are also to be expected in the first 6 month of the life of FAST, given the new discovery space opened up by the large sky coverage and the leap in collecting area realized by FAST.

**Acknowledgments**

This work is supported by the Chinese Academy of Sciences and a key project grant 10433020 from the National Natural Science Foundation in China. We wish to make special recognition to the diligence works and great efforts by colleagues in FAST team and international collaborators, and would also like to apologize for the incomplete coverage of their great contribution to project FAST. This work was carried out in part at the Jet Propulsion Laboratory, California Institute of Technology, under contract with the National Aeronautics and Space Administration.